\documentclass[12pt,journal,a4paper,draftcls,oneside,onecolumn]{IEEEtran}

\usepackage{times} 
\usepackage{graphicx}
\usepackage{cite}
\usepackage{citesort}
\usepackage{color}
\usepackage{psfrag}
\usepackage{subfigure}
\usepackage{amssymb}
\usepackage{epsfig}
\usepackage{pifont}
\usepackage{amsmath}
\usepackage{array}
\usepackage{multicol}
\usepackage[T1]{fontenc}

\renewcommand{\baselinestretch}{1.2} 
\makeatletter
\def\ifundefined{\@ifundefined}
\makeatother 

\begin{document}


\title{{Orthogonal Time-Frequency-Space (OTFS) and Related Signaling}
\author{Lie-Liang Yang}
\thanks{L.-L. Yang is with the School of Electronics and Computer Science, University of Southampton, SO17 1BJ, UK. (E-mail: lly@ecs.soton.ac.uk). This document is a section from book: L.-L. Yang, J. Shi, K.-T. Feng, L.-H. Shen, S.-H. Wu and T.-S. Lee, Resource-Allocation in Wireless Communications: Fundamentals, Algorithms and Applications, Academic Press, USA (to be published in 2024).}}


\maketitle

\begin{abstract}
The principle of orthogonal time-frequency-space (OTFS) signaling is
firstly analyzed, followed by explaining that OTFS embeds another
signaling scheme referred to as orthogonal short-time Fourier
(OSTF). Then, the relationship among OTFS, OSTF, orthogonal
frequency-division multiplexing (OFDM) and single-carrier
frequency-division multiple-access (SC-FDMA) is explored,
demonstrating that OSTF/OTFS are fundamentally the extensions of
OFDM/SC-FDMA from one-dimensional (1D) signaling to two-dimensional
(2D) signaling. Hence, the characteristics and performance of
OSTF/OTFS schemes can be perceived from the well-understood
OFDM/SC-FDMA schemes. Accordingly, the advantages and disadvantages of
OSTF/OTFS are discussed. Furthermore, from the principles of
OFDM/SC-FDMA, the multiuser multiplexing in OSTF/OTFS systems is
analyzed with respect to uplink and downlink, respectively. Added on
this, a range of generalized multiplexing schemes are presented, whose
characteristics are briefly analyzed.
\end{abstract}
%

\section{Introduction}\label{Section-topic-introduction}

In Chapter~1, the principle of OFDM has been analyzed, showing that
OFDM has a range of advantages. With the aid of a suitable cyclic
prefix (CP)\index{Cyclic prefix (CP)}, OFDM is free from inter-symbol
interference (ISI)\index{Inter-symbol interference (ISI)}, when
communicating over frequency-selective fading channels. In OFDM,
one-tap frequency (F)-domain equalization is enough to remove the
effect of channel fading, making it desirably low-complexity for
implementation. In OFDM, the frequency-selective fading channel is
converted to a set of flat-fading channels, which may experience
correlated or independent fading. Owing to this, if transmitter is
designed to employ channel state information (CSI), it can dynamically
load different subcarriers with different amount of information
according to their channel states, to increase system's
spectral-efficiency. Furthermore, power-allocation can be optimized
across subcarriers according to their channel states, so as to meet
different design objectives, as shown in Chapter~3. In multiuser OFDM,
referred to as orthogonal frequency-division multiple-access (OFDMA),
systems, independently faded channels are highly feasible for the
implementation of resource-allocation, including both subcarrier- and
power-allocation, among users, which enables to achieve significant
multiuser diversity gain, as demonstrated in Chapter~4. Owing to the
above-mentioned merits, OFDM became the dominate signaling scheme in
both 4G and 5G standards.

OFDM also has its disadvantages. As analyzed in Chapter~1, it has the
PAPR\index{PAPR} problem, which imposes extra demand on transceiver
design. OFDM signal may experience inter-carrier interference (ICI),
if transmitter/receiver's oscillators are not sufficiently
synchronized, or if channel experiences Doppler-spread due to the
mobilities of transmitter, receiver or/and communication
environments. OFDM needs the help of CP to mitigate ISI, which
penalizes system's spectral-efficiency. When communicating over
time-invariant frequency-selective fading channels, it can be shown
that ICI is mild and the penalty of adding CP is light. However,
wireless channels may be highly time-varying. This is especially the
case when considering the future wireless networks, where various
high-speed transportation systems, satellite networks, etc., are
integrated to support global seamless coverage. Moreover, future
wireless systems are expected to be operated over a widely spanning
bands, from sub-6G to millimeter wave (mmWave) to Terahertz (THz)
bands. Consequently, wireless channels may experience severe
Doppler-spread, resulting in the highly time-varying channels, which
generate big impacts on the transceiver design and performance of
wireless systems.

Specifically, in high Doppler-spread communication environments, the
ICI in OFDM systems becomes severe, which demands a high-complexity
equalizer to mitigate the resulted effect. While this is a problem, an
even worse one may be from the required CP. In OFDM transmission, at
least each OFDM symbol is required to be time-invariant, i.e., each
OFDM symbol's duration should be less than channel's coherence
time\index{Channel coherence bandwidth}, which is the reciprocal of
two timing the maximum Doppler-spread. Hence, when a channel becomes
more time-variant, its coherence time becomes shorter, which in turn
requires to shorten the duration of OFDM symbols. Consequently, given
that channel's maximum delay-spread is fixed and hence the CP length
is fixed, the duration for information transmission within one OFDM
symbol period then has to be shortened. Explicitly, this increases the
overhead for CP and hence reduces system's spectral-efficiency, which
can be significant in fast time-varying environments. Moreover, it also has
the following consequences. To reduce the duration of information
transmission, if system bandwidth keeps fixed, the number of
subcarriers has to be reduced to protect their
orthogonality. Consequently, individual subband's bandwidth needs to
be increased, which is vulnerable to frequency-selective
fading. Alternatively, if the number of subcarriers is fixed,
individual subband's bandwidth has to be increased, and also system's
bandwidth. In this case, the orthogonality of subcarriers cannot be
guaranteed, resulting in the increased ICI, especially, when there is
also big Doppler-spread.

It can be understood that OFDM is an one-dimensional (1D) signaling
scheme, using mainly signal processing in F-domain to conquer the
frequency-selective fading resulted from channel's delay-spread. When
wireless channel conflicts simultaneously both frequency-selective
fading and time-selective fading (resulted from Doppler-spread), the 1D
signaling schemes operated in either the conventional time (T)-domain
or F-domain (like OFDM) present deficiencies. Instead,
two-dimensional (2D) signaling schemes are desirable to simultaneously
deal with the effects generated by channel's delay- and
Doppler-spread. With this motivation, this section considers the 2D
signaling schemes, including the orthogonal short-time Fourier
(OSTF)~\cite{9685566,1347350} and orthogonal time-frequency-space
(OTFS)~\cite{7925924,hadani2018orthogonal} schemes. We will show that
OSTF represents the extension of OFDM, while OTFS is the extension of
the OFDM's companion scheme, namely single-carrier frequency-division
multiple-access (SC-FDMA). Hence, from OFDM and SC-FDMA and their
relationship, we can gain knowledge about the characteristics and
performance of OSTF and OTFS. Furthermore, we can be inferred the
feasibility of OSTF and OTFS for the implementation of
resource-allocation, and which resource-allocation schemes provided in
previous chapters may be extended for application in OSTF and OTFS
systems. Additionally, in this section, the generalizations of OSTF
and OTFS are considered and a range of related schemes for
uplink/downlink multiuser scenarios are presented and briefly
analyzed.

\section{Principles of Orthogonal Time-Frequency-Space (OTFS)}\label{subsection-6G-3.1}

%
\begin{figure}[tb]
  \begin{center}
    \includegraphics[width=0.9\linewidth]{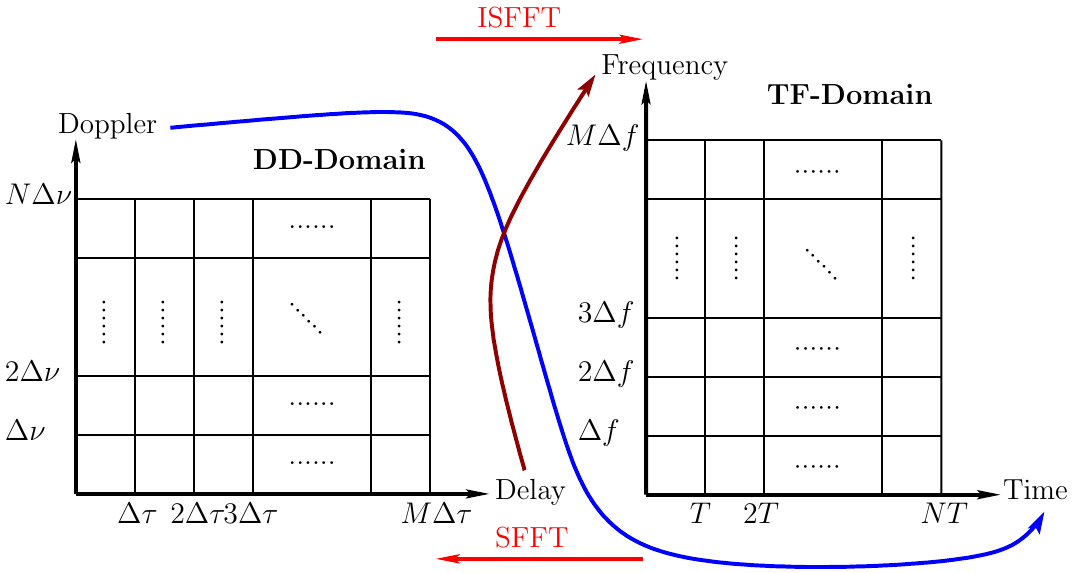}
 
  \end{center}
  \caption{OTFS signaling relationship between DD-domain and TF-domain.}
\label{figure-DD-TF-relationship}
\end{figure}

Assume a signal-antenna transmitter sending information to a
single-antenna receiver for simplicity to focus on the principles of
OTFS. Assume that information symbols are sent in blocks of each block
consisting of $MN$ symbols over a channel having the total bandwidth
$B$. The block-duration is expressed as $T_B$. To facilitate OTFS
signaling, the bandwidth $B$ is divided into $M$ subbands of each
having a bandwidth $\Delta f=B/M$; the block-duration $T_B$ is divided
into $N$ time-slots of each with a duration of $T=T_B/N$, forming the
time-frequency (TF)-grid\index{Time-frequency (TF)-grid} as shown at
the right-side of Fig.~\ref{figure-DD-TF-relationship}. It is assumed
that $T\Delta f=1$ to protect the
orthogonality~\cite{7925924,8424569}. With these settings, it is
known that the delay resolution and Doppler resolution of the OTFS
signals are $\Delta\tau=1/B=1/M\Delta f$ and $\Delta\nu=1/T_B=1/NT$,
respectively~\cite{COM-Bello-1963}.

Assume that the channel's maximum delay-spread\index{Delay-spread}
satisfies $\tau_{\max}\leq T$ and the channel's maximum
Doppler-spread\index{Doppler-spread} is limited to $|\nu_{\max}|\leq
\Delta f/2$. Corresponding to the TF-grid, using the delay and Doppler
resolutions, the delay-Doppler (DD)-plan in the region of
$\left(\tau\in[0,T]\times \nu\in[-\Delta f/2,\Delta f/2]\right)$ is
divided into the grid as shown at the left-side of
Fig.~\ref{figure-DD-TF-relationship}, where on the delay-axis, each
division accounts for a delay-spread difference of $\Delta \tau$ and
on the Doppler-axis, each division represents a Doppler-spread difference of
$\Delta \nu$. In Fig.~\ref{figure-DD-TF-relationship}, the
relationships between the delay-spread/Doppler-spread in delay-Doppler
(DD)-domain and the subbands/time-slots in time-frequency (TF)-domain
are explicitly illustrated. For clarity, in the analysis below, the
$m$-related indices, i.e., $\tilde m$ and $m$, are used for
delay-spread and subbands, while the $n$-related indices, i.e.,
$\tilde n$ and $n$, are used for Doppler-spread and
time-slots. $\tilde x[~~]$ and $\tilde y[~~]$ are in DD-domain, and
$X[~~]$ and $Y[~~]$ are in TF-domain.

\begin{figure}[tb]
  \begin{center}
    \includegraphics[width=0.8\linewidth]{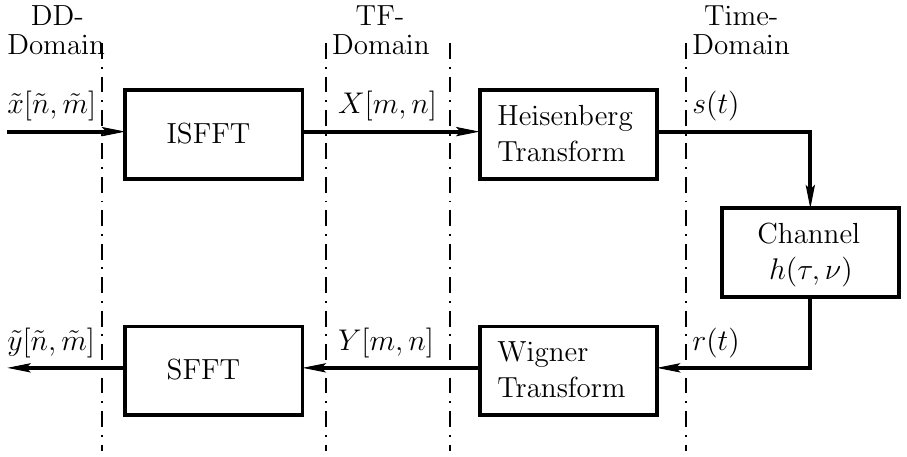}
 
  \end{center}
  \caption{Modulation and demodulation signaling in OTFS principles.}
\label{figure-OTFS-diagram}
\end{figure}

Following \cite{7925924,hadani2018orthogonal,8424569,9321356}, the
principles of OTFS can be summarized as follows. Assume that a block
of $MN$ data symbols, such as QAM/PSK symbols, are to be
transmitted. These symbols are mapped to a $(N\times M)$ grid, as
shown by the left-side one in Fig.~\ref{figure-DD-TF-relationship},
corresponding to a $(N\times M)$-dimensional matrix $\tilde{\pmb{X}}$,
with the element on row $\tilde{n}$ and column $\tilde{m}$ expressed
as $\tilde x[\tilde n, \tilde m]$. As shown in
Fig.~\ref{figure-OTFS-diagram}, $\tilde{\pmb{X}}$ is mapped to a
$(M\times N)$-dimensional matrix $\pmb{X}$ in TF-domain via the
inverse symplectic finite Fourier transform
(ISFFT)~\cite{7925924,hadani2018orthogonal}\index{Transform ! inverse symplectic
  finite Fourier transform (ISFFT)}\index{Transform ! Symplectic finite Fourier
  transform (SFFT)}, with the $(m,n)$-th element expressed as
\begin{align}\label{eq:OTFS-1}
X[m,n]=&\frac{1}{\sqrt{MN}}\sum_{\tilde m=0}^{M-1}\sum_{\tilde n=0}^{N-1}\tilde{x}[\tilde n,\tilde m]\exp\left(j2\pi\left[\frac{n\tilde n}{N}-\frac{m\tilde m}{M}\right]\right)\nonumber\\
&~~m=0,1,\ldots,M-1;~~n=0,1,\ldots,N-1
\end{align}    
It can be shown that in matrix form, we have 
\begin{align}\label{eq:OTFS-2}
\pmb{X}=\pmb{\mathcal{F}}_{M}\tilde{\pmb X}^T\pmb{\mathcal{F}}^H_{N}
\end{align}
where $\pmb{\mathcal{F}}$ are the normalized FFT matrix defined in (1.15) in Chapter~1.

As shown in Fig.~\ref{figure-OTFS-diagram}, then, the Heisenberg
transform\index{Transform ! Heisenberg transform}, also referred to as
orthogonal short-time Fourier (OSTF)
transform~\cite{Sayeed-TF-perspective,9685566,1347350,Cohen-Time-Frequency-book}\index{Transform
  ! orthogonal short-time Fourier (OSTF)}, is executed to transform
the signals from TF-domain to T-domain for transmission, which
generates the T-domain transmit signal as
\begin{align}\label{eq:OTFS-3}
s(t)=\sum_{n=0}^{N-1}\sum_{m=0}^{M-1}X[m,n]g_{tx}(t-nT)\exp\left(j2\pi m\Delta f[t-nT]\right)
\end{align}
In \eqref{eq:OTFS-3}, $g_{tx}(t)$ is the pulse-shaping filter of TF
modulation, which is defined in $(0,T]$ and normalized to have unit
  power, i.e., $T^{-1}\int_0^T|g(t)|^2dt=1$. Furthermore, the transmit
  pulse $g_{tx}(t)$ and its corresponding receive pulse $g_{rx}(t)$
  satisfy the bi-orthogonality\index{Bi-orthogonality} expressed
  as~\cite{7925924,hadani2018orthogonal}
\begin{align}\label{eq:OTFS-4}
\int_{nT}^{(n+1)T} &g^*_{rx}(t-n'T)\exp\left(-j2\pi m'\Delta f[t-n'T]\right)g_{tx}(t-nT)\nonumber\\
&\times\exp\left(j2\pi m\Delta f[t-nT]\right)dt=\delta[n-n']\delta[m-m']
\end{align}
where $\delta[\cdot]$ is the Dirac delta function. 

From the TF-grid, we know that the TF resources can provide
\begin{align}\label{eq:OTFS-5}
\frac{T_B}{\frac{1}{B}}=\frac{NT}{\frac{1}{M\Delta f}}=MNT\Delta f=MN
\end{align} 
degrees of freedom (DoFs). Hence, there are $MN$ independent basis waveforms available. Correspondingly, \eqref{eq:OTFS-3} explains that the $MN$ elements of $\pmb{X}$ are modulated onto the $MN$ OSTF basis waveforms given by~\cite{9685566}
\begin{align}\label{eq:OTFS-6}
\phi_{m,n}(t)=&g_{tx}(t-nT)\exp\left(j2\pi m\Delta f[t-nT]\right)\nonumber\\
&~~m=0,1,\ldots,M-1;~~n=0,1,\ldots,N-1
\end{align} 

Assume that a time-varying channel is characterized by the
complex baseband impulse response $h(\tau,\nu)$ in the DD-domain. When
$s(t)$ of \eqref{eq:OTFS-3} is transmitted over this channel, the
received signal is then given by
\begin{align}\label{eq:OTFS-7}
r(t)=\int\int h(\tau,\nu)s(t-\tau)e^{j2\pi \nu(t-\tau)}d\tau d\nu+n(t)
\end{align}
where $n(t)$ is Gaussian noise.

At receiver, as shown in Fig.~\ref{figure-OTFS-diagram}, $r(t)$ is
transformed to TF-domain using the Wigner transform\index{Transform !
  Wigner}, which implements the 2D matched-filtering to compute the
ambiguity function\index{Ambiguity function}
as~\cite{Cohen-Time-Frequency-book}
\begin{align}\label{eq:OTFS-8}
Y[m,n]=\int_{nT}^{(n+1)T} r(t)g_{rx}^*(t-nT)\exp\left(-j2\pi m\Delta f[t-nT]\right)dt
\end{align}
It can be shown that $Y[m,n]$ can be expressed as~\cite{8424569}
\begin{align}\label{eq:OTFS-9}
Y[m,n]=\sum_{m'=0}^{M-1}\sum_{n'=0}^{N-1}H_{m,n}[m',n']X[m',n']+N[m,n]
\end{align}
where 
\begin{align}\label{eq:OTFS-10}
N[m,n]=\int_{nT}^{(n+1)T} n(t)g_{rx}^*(t-nT)\exp\left(-j2\pi m\Delta f[t-nT]\right)dt
\end{align}
and
\begin{align}\label{eq:OTFS-11}
H_{m,n}&[m',n']=\int\int h(\tau,\nu)A_{g_{rx},g_{tx}}[(n-n')T-\tau),(m-m')\Delta f-\nu]\nonumber\\
&\times \exp\left(j2\pi[v+m'\Delta f][(n-n')T-\tau]\right)\exp\left(j2\pi \nu n'T\right)d\tau d\nu
\end{align}

Note that, $H_{m,n}[m',n']$ is in fact the coupling generated by the channel between the transmit basis function $\phi_{m',n'}(t)$ and the receive basis function $\phi_{m,n}(t)$, represented as~\cite{1347350}
\begin{align}\label{eq:OTFS-12}
H_{m,n}&[m',n']=\int H(\phi_{m',n'}(t))\phi_{m,n}^*(t)dt
\end{align}
with 
\begin{align}\label{eq:OTFS-13}
H(\phi_{m',n'}(t))=\int\int h(\tau,\nu)\phi_{m',n'}(t-\tau)e^{j2\pi \nu(t-n'T-\tau)}d\tau d\nu
\end{align}

For simplicity, assume that the ambiguity function seen in \eqref{eq:OTFS-11} satisfies 
\begin{align}\label{eq:OTFS-14}
A_{g_{rx},g_{tx}}[(n-n')T-\tau_{\max}),(m-m')\Delta f-\nu_{\max}]\approx\delta[n-n']\delta[m-m']
\end{align}
where $\tau_{\max}$ and $\nu_{\max}$ denote the expected maximum
delay-spread and Doppler-spread, respectively. Then, the TF-domain
channel $H[m,n]$ in \eqref{eq:OTFS-11} can be expressed as
\begin{align}\label{eq:OTFS-15}
H[m,n]=&H_{m,n}[m,n]\nonumber\\
=& \int\int \exp\left(-j2\pi v\tau\right) h(\tau,\nu) \exp\left(-j2\pi[m\Delta f\tau-n\nu T]\right)d\tau d\nu
\end{align}
When sampling it in DD-domain using $\Delta \tau=1/(M\Delta f)$ and $\Delta\nu=1/(NT)$, we obtain
\begin{align}\label{eq:OTFS-16}
H[m,n]=& \sum_{m'=0}^{M-1}\sum_{n'=0}^{N-1}  h\left(m',n'\right)\exp\left(-j2\pi\left[\frac{mm'}{M}-\frac{nn'}{N}\right]\right)\nonumber\\
&\times \exp\left(-j 2\pi\frac{ m'n'}{MN}\right)
\end{align}
where $h(m',n')=h\left(\frac{m'}{M\Delta
  f},\frac{n'}{NT}\right)$. Assume that $h(m',n')\neq 0$ only when
$m'\leq L_{\max}$ and $n'\leq V_{\max}$, where $L_{\max}$ represents
the maximum number of resolvable paths\index{Resolvable paths !
  delay-domain} in delay (D)-domain, while $V_{\max}$ represents the
maximum number of resolvable paths in Doppler-domain\index{Resolvable
  paths !  Doppler-domain}. Then, $h(m',n')$ with $m'=0,1,\ldots,
L_{\max}-1$ and $n'=0,1,\ldots,V_{\max}-1$ gives the channel's impulse
respond in DD-domain. Correspondingly, \eqref{eq:OTFS-16} is
\begin{align}\label{eq:OTFS-17}
H[m,n]=& \sum_{m'=0}^{L_{\max}-1}\sum_{n'=0}^{V_{\max}-1}  h\left(m',n'\right)\exp\left(-j2\pi\left[\frac{mm'}{M}-\frac{nn'}{N}\right]\right)\nonumber\\
&\times \exp\left(-j 2\pi\frac{ m'n'}{MN}\right)\\
\label{eq:OTFS-17a}
=& \sum_{m'=0}^{L_{\max}-1}\sum_{n'=0}^{V_{\max}-1}  \tilde{h}\left(m',n'\right)\exp\left(-j2\pi\left[\frac{mm'}{M}-\frac{nn'}{N}\right]\right)
\end{align}
where $\tilde{h}\left(m',n'\right)=h\left(m',n'\right)\exp\left(-j
2\pi\frac{ m'n'}{MN}\right)$. Let $\pmb{H}_{TF}=\{H[m,n]\}$,
$\pmb{H}_{DD}=\{h[m,n]\}$ and its modified version
$\tilde{\pmb{H}}_{DD}=\{\tilde{h}[m,n]\}$. Explicitly,
$\tilde{\pmb{H}}_{DD}$ is known once $\pmb{H}_{DD}$ is known, or vice
versa. When $\tilde{\pmb{H}}_{DD}$ is a $(L_{\max}\times V_{\max})$
matrix containing only the elements indexed by $m'$ and $n'$ in
\eqref{eq:OTFS-17a}, we have
\begin{align}\label{eq:OTFS-17b}
\pmb{H}_{TF}=&\sqrt{MN}\pmb{\mathcal{F}}_M\pmb{\Psi}_{M\times
  L_{\max}}\tilde{\pmb{H}}_{DD}\pmb{\Psi}^T_{N\times
  V_{\max}}\pmb{\mathcal{F}}_N^H\nonumber\\ 
=&\sqrt{MN}\pmb{\mathcal{F}}_M\tilde{\pmb{H}}_{DD}^{(s)}\pmb{\mathcal{F}}_N^H
\end{align}
where $\pmb{\Psi}_{M\times L_{\max}}$ is a mapping matrix,
constituting the $L_{\max}$ columns of $\pmb{I}_M$ corresponding to
$m'$ seen in \eqref{eq:OTFS-17a}, and similarly, $\pmb{\Psi}_{N\times
  V_{\max}}$ is constructed. In \eqref{eq:OTFS-17b},
$\tilde{\pmb{H}}_{DD}^{(s)}=\pmb{\Psi}_{M\times
  L_{\max}}\tilde{\pmb{H}}_{DD}\pmb{\Psi}^T_{N\times V_{\max}}$ is a
$(M\times N)$ sparse matrix, if there are only a small number of paths
in DD domain, when compared with $MN$. In this case, it benefits the
estimation of TF-selective fading channels in DD-domain with the aid
of compressed sensing algorithms~\cite{book:Zhang-Xianda}. 

The above analysis and \eqref{eq:OTFS-17b} explain that
$\pmb{H}_{TF}$, $\pmb{H}_{DD}$ and $\tilde{\pmb{H}}_{DD}$ are
equivalent, provided that the ranks of the matrices on the left and
right sides of \eqref{eq:OTFS-17b} can satisfy the requirements of
$M\geq L_{\max}$ and $N\geq V_{\max}$. When one of them is known,
the other two can be mathematically derived. Below are two special
forms.

First, when channel only experiences delay-spread resulted
frequency-selective fading but no Doppler-spread resulted
time-selective fading, we have $N=1$ and $n'=0$. In this case,
\eqref{eq:OTFS-17} is reduced to
\begin{align}\label{eq:OTFS-18}
H[m]=& \sum_{m'=0}^{L_{\max}-1} h\left(m'\right)\exp\left(-j2\pi\left[\frac{mm'}{M}\right]\right),~m=0,1,\ldots,M-1
\end{align}
Accordingly, $H[m]$ is the fading gain of the $m$th subcarrier of an
OFDM system with $M$ subcarriers, as seen in Section~1.5 of
Chapter~1. When $L_{\max}\leq M/2$, the $M$ subcarrier channels exist
correlation, as analyzed in Section~1.5. Corresponding to
\eqref{eq:OTFS-17b},
$\pmb{h}_{F}=\sqrt{M}\pmb{\mathcal{F}}_M\pmb{\Psi}_{M\times
  L_{\max}}{\pmb{h}}_{D}$ builds the fading relationship between
D-domain and F-domain, where $\pmb{h}_{F}$ and ${\pmb{h}}_{D}$ are
column vectors, and ${\pmb{h}}_{D}$ is the channel impulse response
(CIR)\index{Channel impulse response (CIR)} in D-domain.

Second, when channel only experiences the Doppler-spread resulted
time-selective fading, but no delay-spread resulted
frequency-selective fading, we have $M=1$ and $m'=0$. In this case,
\eqref{eq:OTFS-17} is simplified to
\begin{align}\label{eq:OTFS-19}
H[n]=& \sum_{n'=0}^{V_{\max}-1}  h\left(n'\right)\exp\left(j2\pi\left[\frac{nn'}{N}\right]\right),~n=0,1,\ldots, N-1
\end{align}
Correspondingly, $H[n]$ is the fading gain experienced by the $n$th
time-slot. Similarly, it can be shown that, when $V_{\max}\leq N/2$,
the $N$ time-slots experience correlated fading. Corresponding to
\eqref{eq:OTFS-17b},
$\pmb{h}_{T}=\pmb{\mathcal{F}}_N^H\pmb{\Psi}_{N\times
  V_{\max}}\pmb{h}_{D}$ gives the fading relationship between
Doppler-domain and T-domain, where $\pmb{h}_{T}$ and ${\pmb{h}}_{D}$
are column vectors, while ${\pmb{h}}_{D}$ is the CIR in
Doppler-domain.

Let us now return to \eqref{eq:OTFS-9}. When the ambiguity function
satisfies the condition of \eqref{eq:OTFS-14}, $H_{m,n}[m',n']$ in
\eqref{eq:OTFS-11} is non-zero only when $m'=m$ and $n'=n$. Hence,
\eqref{eq:OTFS-9} is simplified to
\begin{align}\label{eq:OTFS-20}
Y[m,n]=H[m,n]X[m,n]+N[m,n],~&m=0,1,\ldots,M-1;\nonumber\\
&n=0,1,\ldots,N-1
\end{align}
showing that there is no interference between the TF-domain symbols
sent in TF-domain, revealing the characteristics of OFDM experiencing
only frequency-selective fading. However, we should be aware of that
this is achieved, only when the condition of \eqref{eq:OTFS-14} is
guaranteed. If this is not the case due to the strong selectivity in
TF-domain or/and the non-ideal $g_{rx}(t)$ and $g_{tx}(t)$, the
detection of TF-domain symbols $X[m,n]$ would experience inter-symbol
interference (ISI)\index{Inter-symbol interference (ISI)}.

Since at transmitter, data symbols are mapped to the DD-domain grid,
as shown in Fig.~\ref{figure-DD-TF-relationship}, at receiver, as
shown in Fig.~\ref{figure-OTFS-diagram}, the TF-domain signals are
converted to the DD-domain using the symplectic finite Fourier
transform (SFFT)\index{Symplectic finite Fourier
transform (SFFT)} as
\begin{align}\label{eq:OTFS-21}
\tilde{y}[\tilde{n},\tilde{m}]=&\frac{1}{\sqrt{MN}}\sum_{m=0}^{M-1}\sum_{n=0}^{N-1}Y[m,n]\exp\left(-j2\pi\left[\frac{\tilde{n}n}{N}-\frac{\tilde{m}m}{M}\right]\right),\nonumber\\
&~~\tilde{n}=0,1,\ldots,N-1;~~\tilde{m}=0,1,\ldots,M-1
\end{align}
Upon substituting \eqref{eq:OTFS-20} associated with \eqref{eq:OTFS-1} and \eqref{eq:OTFS-15} into \eqref{eq:OTFS-21}, it can be shown that the input-output relationship in DD-domain can be represented as~\cite{8424569,7925924,hadani2018orthogonal}
\begin{align}\label{eq:OTFS-22}
\tilde{y}[\tilde{n},\tilde{m}]=&\frac{1}{MN}\sum_{\tilde{m}'=0}^{M-1}\sum_{\tilde{n}'=0}^{N-1}\tilde{x}[\tilde{n},\tilde{m}]h_w[\tilde{m}-\tilde{m}',\tilde{n}-\tilde{n}']+\tilde{n}[\tilde{n},\tilde{m}]
\end{align}
where 
\begin{align}\label{eq:OTFS-23}
\tilde{n}[\tilde{n},\tilde{m}]=&\frac{1}{\sqrt{MN}}\sum_{m=0}^{M-1}\sum_{n=0}^{N-1}N[m,n]\exp\left(-j2\pi\left[\frac{\tilde{n}n}{N}-\frac{\tilde{m}m}{M}\right]\right)
\end{align}
is the Gaussian noise, $h_w[l,k]$ is the discrete representation of the
DD-domain channel impulse response $h_w(\tau,\nu)$ upon considering
the effect from the rectangular windowing functions applied at both
transmitter and receiver\footnote{Note that, for simplicity, the
  windowing functions were not explicitly shown, such as, in
  \eqref{eq:OTFS-3} and \eqref{eq:OTFS-21}. Otherwise, $X[m,n]$ and
  $Y[m,n]$ should be replaced by $W_{tx}[m,n]X[m,n]$ and
  $W_{rx}[m,n]Y[m,n]$, respectively, where $W_{tx}[m,n]$ and
  $W_{rx}[m,n]$ are the windowning functions in TF-domain.}, which can
be expressed as
\begin{align}\label{eq:OTFS-24}
h_w(\tau,\nu)=\int\int h(\tau',\nu')w(\tau-\tau',\nu-\nu')\exp(-j2\pi \nu\tau)d\tau' d\nu'
\end{align}
with
\begin{align}\label{eq:OTFS-25}
w(\tau,\nu)=\sum_{m=0}^{M-1}\sum_{n=0}^{N-1}\exp\left(-j2\pi [\nu nT-\tau m\Delta f]\right)
\end{align}
Hence,
\begin{align}\label{eq:OTFS-26}
h_w[l,k]=h_w\left(\frac{l}{M\Delta f},\frac{k}{NT}\right)
\end{align}

Note that, upon discretizing the variables in \eqref{eq:OTFS-24}, we
have an expression
\begin{align}\label{eq:OTFS-27}
h_w(l,k)=\sum_{l'=0}^{M-1}\sum_{k'=0}^{N-1} h(l',k')w(l-l',k-k')\exp\left(-j\frac{2\pi lk}{MN}\right)
\end{align}

The input-output relationship in \eqref{eq:OTFS-22} explains that the
observation $\tilde{y}[\tilde{n},\tilde{m}]$ corresponding to
$\tilde{x}[\tilde{n},\tilde{m}]$ consists of a linear combination of
possibly all the transmitted symbols
$\tilde{x}[\tilde{n}',\tilde{m}']$. Hence, there is in general ISI,
which needs to be mitigated in detection by a suitable equalizer, such
as, that is implemented in the principle of minimum mean-square error
(MMSE)\index{Minimum mean-square error}~\cite{9685566}.

\section{Relationship of OTFS with OSTF, OFDM and SC-FDMA}\label{subsection-6G-3.2}

Now we analyze the relationship of OTFS with some other multicarrier
signaling schemes, including the orthogonal short-time Fourier
(OSTF)~\cite{9685566,1347350}, OFDM and single-carrier
frequency-division multiple-access
(SC-FDMA)~\cite{Lie-Liang-MC-CDMA-book} schemes. From their
relationship, the characteristics of OTFS and OSTF as well as their
advantages and disadvantages in applications will become clearer based
on our knowledge about SC-FDMA and OFDM. Please refer to
\cite{9685566,1347350} for the details of OSTF signaling, the
principle of OFDM has been addressed in Chapter~1, while the principle
of SC-FDMA can be found in \cite{Lie-Liang-MC-CDMA-book}, all of which
are also inferred in the following texts.

First, according to \cite{9685566,1347350}, the OSTF signaling is
embedded in the OTFS signaling. It has a system diagram as shown in
Fig.~\ref{figure-OTFS-diagram} after removing the ISFFT at transmitter
and SFFT at receiver. Therefore, in OSTF signaling, information
symbols are mapped on a $(M\times N)$-grid in TF-domain, and the
transmit signal is given by \eqref{eq:OTFS-3}, where $X[m,n]$ are data
symbols. Alternatively, the OTFS signaling in \eqref{eq:OTFS-3} is
reduced to the OSTF signaling, when $\pmb{\mathcal{F}_M}$ and
$\pmb{\mathcal{F}_N}$ are replaced by the identity matrices of
corresponding dimensions.

\begin{figure}[tb]
  \begin{center}
    \includegraphics[width=0.7\linewidth]{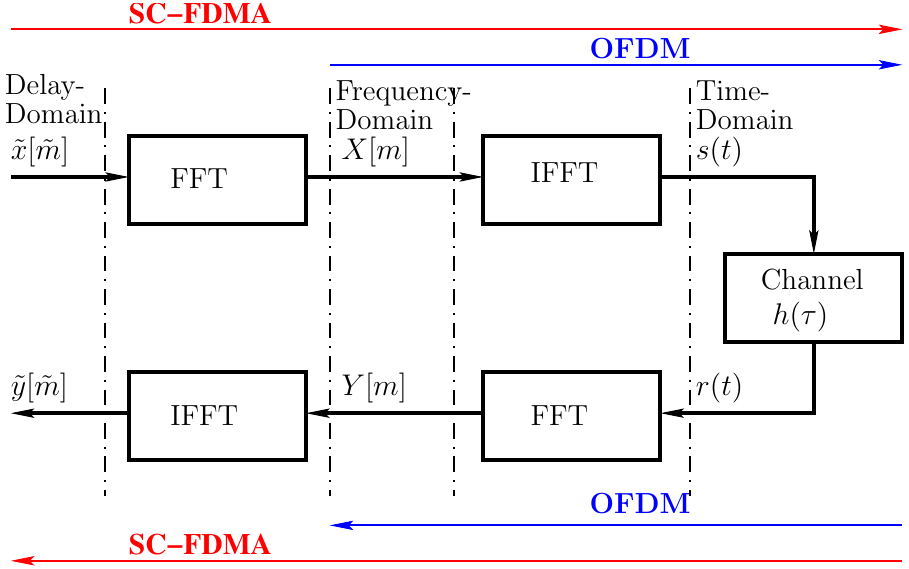}
 
  \end{center}
  \caption{Diagram illustrating the relationship between OFDM and SC-FDMA signaling.}
\label{figure-OFDM-SC-FDMA-diagram}
\end{figure}

Second, when $N=1$ and assuming that $g_{tx}(t)$ is a rectangular
waveform function having unity amplitude,  \eqref{eq:OTFS-3}
becomes
\begin{align}\label{eq:OTFS-28}
s(t)=\sum_{m=0}^{M-1}X[m]\exp\left(j2\pi m\Delta ft\right)
\end{align}
which is the OFDM signaling. Also, Eq.~\eqref{eq:OTFS-18} shows that,
when channel has no Doppler-spread, the fading gains of subcarrier channels are
given by the FFT on the CIR in D-domain, which agrees with that in
OFDM systems, as shown in Section~1.5 of
Chapter~1. Hence, OSTF represents the 2D extension of
OFDM.

Third, The relationship between OFDM and SC-FDMA is depicted in
Fig.~\ref{figure-OFDM-SC-FDMA-diagram}. Hence, in SC-FDMA systems,
$\{X[m]\}$ in \eqref{eq:OTFS-28} are provided by the DFT operating on
the data symbols $\{\tilde{x}[\tilde{m}]\}$, i.e.,
$\pmb{X}=\pmb{\mathcal{F}}\tilde{\pmb{x}}$. Hence, based on the
principles of OTFS, we can view that the data symbols in SC-FDMA are
mapped to a vector in D-domain. 

It is worth noting that, in point-to-pint communications where all
resources in TF-domain are assigned to one transmitter, the FFT/IDFT
seen in the transmitter of Fig.~\ref{figure-OFDM-SC-FDMA-diagram} is
meaningless, as $\pmb{\mathcal{F}}\pmb{\mathcal{F}}^H=\pmb{I}$. In
this case, data symbols $\{\tilde{x}[\tilde{m}]\}$ are directly
transmitted in blocks in T-domain, with each block adding a suitable
CP to mitigate inter-block interference. In this way, at receiver,
after removing the CP and transforming the received signals to F-domain
using a FFT, the low-complexity one-tap F-domain
equalization\index{Frequency-domain equalization} can be carried out
to mitigate ISI~\cite{J-Adachi-IEICE-2009-1441,5493922}, before
transforming the signals to D-domain to carry out data detection. In
practice, SC-FDMA scheme is typically employed to support the uplink
multiuser transmissions~\cite{Lie-Liang-MC-CDMA-book}. In this regard,
each user is assigned a portion of $M_d=M/K$ subcarriers, where $K$
denotes the number of users. Correspondingly, at each user's
transmitter, the FFT matrix is $(M_d\times M_d)$ dimensional. After
FFT, the $M_d$ symbols are mapped to $M_d$ out of $M$ subcarriers
based on a mapping scheme. Specifically, interleaved mapping, giving
the so-called interleaved frequency-division multiple-access
(IFDMA)\index{Interleaved frequency-division multiple-access (IFDMA)},
or localized mapping, yielding the localized FDMA
(LFDMA)\index{Localized frequency-division multiple-access (IFDMA)},
can be implemented. More generally, many other, including random,
mapping may also be implemented. The selection between IFDMA and LFDMA
is depended on the design objective of attaining frequency diversity
gain or benefit from resource-allocation, which will be further
discussed later in conjunction with OTFS/OSTF.

Based on the relationship between OFDM and SC-FDMA and that between
OFDM and OSTF, it is easy to show that OTFS is reduced to SC-FDMA,
when $N=1$ and assuming that $g_{tx}(t)$ is rectangular with unity
amplitude. This can be seen from \eqref{eq:OTFS-1} and
\eqref{eq:OTFS-2}. Specifically, when $N=1$, \eqref{eq:OTFS-2} becomes
$\pmb{X}=\pmb{\mathcal{F}}\tilde{\pmb{x}}$, which is the FFT operation
implemented in the SC-FDMA transmitter, as shown in
Fig.~\ref{figure-OFDM-SC-FDMA-diagram}. Hence, OTFS represents the 2D
extension of SC-FDMA.

\begin{figure}[tb]
  \begin{center}
    \includegraphics[width=0.7\linewidth]{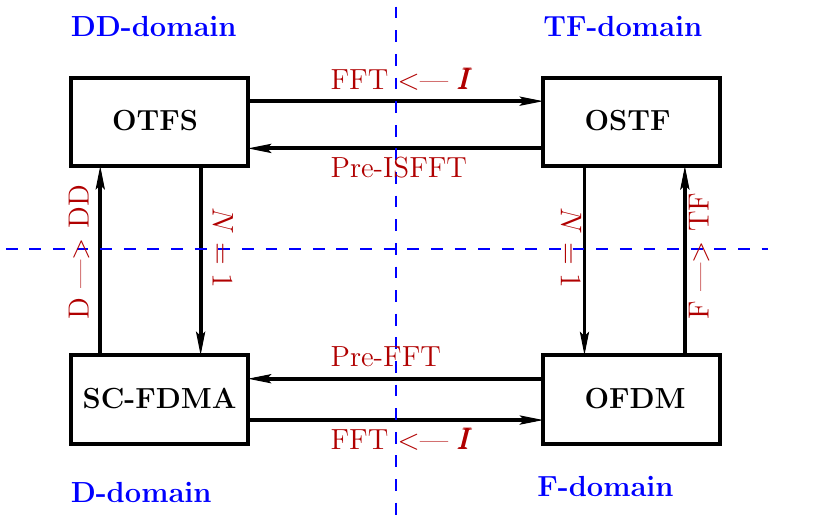}
 
  \end{center}
  \caption{Diagram illustrating the relationship among OTFS, OSTF, OFDM and SC-FDMA.}
\label{figure-OTFS-other-relationship}
\end{figure}

In summary, the relationship among OTFS, OSTF, OFDM and SC-FDMA can be
depicted as Fig.~\ref{figure-OTFS-other-relationship}, where FFF
$\leftarrow\pmb{I}$ means that FFT matrix is replaced by identity
matrix, D$\rightarrow$DD and F$\rightarrow$TF denote the
generalization from D-domain to DD-domain and from F-domain to
TF-domain, respectively, `Pre-FFT' means that the inputs to OFDM are
preprocessed by FFT, and `Pre-ISFFT' indicates that the inputs to OSTF
are preprocessed using ISFFT.

\section{Multiuser Multiplexing and Resource-Allocation in OTFS and OSTF Systems}\label{subsection-6G-3.3}

Having shown the mirror relationship between OTFS/OSTF and SC-FDMA/OFDM
in Section~\ref{subsection-6G-3.2}, let us now analyze the multiuser
multiplexing\index{Multiuser multiplexing} and
resource-allocation\index{Resource-allocation} in OTFS and OSTF
systems. For convenience of description, assume $M_d=M/K_d$ and
$N_D=N/K_D$, where $M_d$, $N_D$, $K_d$ and $K_D$ are all
integers. Then, a block of resources provided by the $(N\times
M)$-grid in DD-domain or provided by the $(M\times N)$-grid in
TF-domain can be simultaneously allocated to $K_dK_D$ users. Each user
can be assigned $M_dN_D$ units of resources or DoFs.

For analysis, the CIR\index{Channel impulse response} in DD-domain is
expressed as
\begin{align}\label{eq:OTFS-29}
h(\tau,\nu)=\sum_{l=0}^{\mathcal{L}}h_l\delta(\tau-\tau_l)\delta(\nu-\nu_{l})
\end{align}
where $\mathcal{L}$ denotes the number of physical propagation paths
(rays), and $h_l$, $\tau_l$ and $\nu_l$ represent the gain, delay and
Doppler-shift of the $l$th path. Often, $l=0$ corresponds to the
line-of-sight (LoS) path\index{Propagation path ! line-of-sight
  (LoS)}, while the others are reflected paths\index{Propagation
  path!reflected}.

In a richly scattered dispersive channel\index{Channel ! richly
  scattered dispersive} operated in relatively low frequency bands,
such as sub-6GHz, $\mathcal{L}$ may be a very big number. When
$\mathcal{L}>>L_{\max}V_{\max}$, where $L_{\max}$ and $V_{\max}$ are
defined in association with \eqref{eq:OTFS-17}, the CIR can be
expressed as
\begin{align}\label{eq:OTFS-30}
h(\tau,\nu)=\sum_{l=0}^{L_{\max}-1}\sum_{d=0}^{V_{\max}-1}h(l,d)\delta\left(\tau-\frac{l}{M\Delta f}\right)\delta\left(\nu-\frac{d}{NT}\right)
\end{align}
explaining that the channel can be resolved into $L_{\max}V_{\max}$
independent paths. Each resolvable path is a combination of many paths
closely related in terms of delay and Doppler-shift.  Each path has a
fading gain of $h(l,d)$, which statistically obeys, such as, complex
Gaussian distribution or other distribution, depended on the
communications environments. The total power conveyed to receiver by
the resolvable paths is
\begin{align}\label{eq:OTFS-31}
P_r=\sum_{l=0}^{L_{\max}-1}\sum_{d=0}^{V_{\max}-1}\|h(l,d)\|^2
\end{align}
which can be exploited by receiver when it coherently combines, such
as, using maximum ratio combining (MRC)\index{Maximum ratio combining
  (MRC)}\cite{Proakis-5th}, all the resolvable paths in DD-domain. It
can be demonstrated that $P_r$ converges to a constant, when
$L_{\max}V_{\max}$ increases.

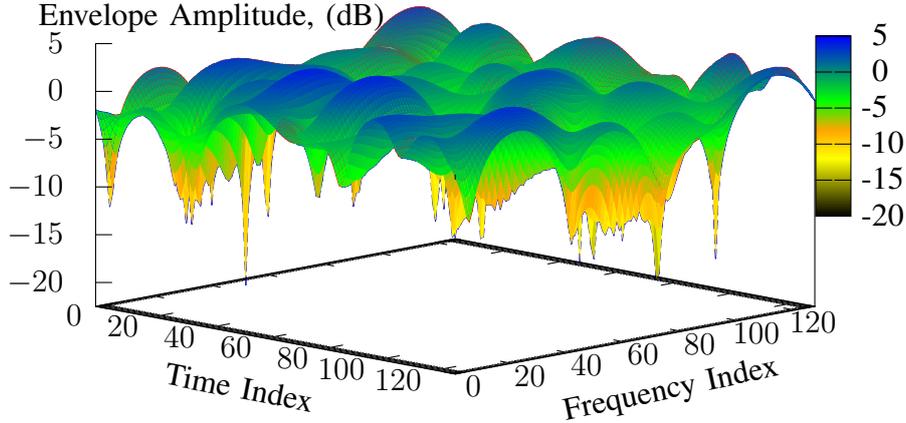
\begin{figure}[t]
  \begin{center}
\begin{picture}(0,0)%
\includegraphics{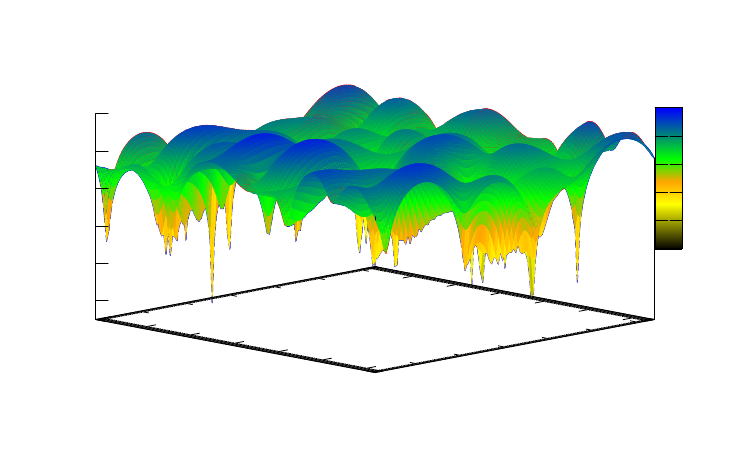}%
\end{picture}%
\begingroup
\setlength{\unitlength}{0.0200bp}%
\begin{picture}(18000,10800)(0,0)%
\put(16579,4827){\makebox(0,0)[l]{\strut{}-20}}%
\put(16579,5504){\makebox(0,0)[l]{\strut{}-15}}%
\put(16579,6181){\makebox(0,0)[l]{\strut{}-10}}%
\put(16579,6858){\makebox(0,0)[l]{\strut{}-5}}%
\put(16579,7535){\makebox(0,0)[l]{\strut{} 0}}%
\put(16579,8212){\makebox(0,0)[l]{\strut{} 5}}%
\put(1236,8587){\makebox(0,0)[l]{\strut{}Envelope Amplitude, (dB)}}%
\put(2038,2934){\makebox(0,0)[r]{\strut{}$0$}}%
\put(3095,2734){\makebox(0,0)[r]{\strut{}$20$}}%
\put(4151,2534){\makebox(0,0)[r]{\strut{}$40$}}%
\put(5207,2334){\makebox(0,0)[r]{\strut{}$60$}}%
\put(6263,2134){\makebox(0,0)[r]{\strut{}$80$}}%
\put(7319,1933){\makebox(0,0)[r]{\strut{}$100$}}%
\put(8376,1733){\makebox(0,0)[r]{\strut{}$120$}}%
\put(4970,1578){\rotatebox{-11}{\makebox(0,0){\strut{}Time Index}}}%
\put(9341,1606){\makebox(0,0){\strut{}$0$}}%
\put(10397,1806){\makebox(0,0){\strut{}$20$}}%
\put(11453,2007){\makebox(0,0){\strut{}$40$}}%
\put(12510,2207){\makebox(0,0){\strut{}$60$}}%
\put(13566,2407){\makebox(0,0){\strut{}$80$}}%
\put(14622,2607){\makebox(0,0){\strut{}$100$}}%
\put(15678,2807){\makebox(0,0){\strut{}$120$}}%
\put(13030,1578){\rotatebox{12}{\makebox(0,0){\strut{}Frequency Index}}}%
\put(1694,3580){\makebox(0,0)[r]{\strut{}$-20$}}%
\put(1694,4478){\makebox(0,0)[r]{\strut{}$-15$}}%
\put(1694,5376){\makebox(0,0)[r]{\strut{}$-10$}}%
\put(1694,6273){\makebox(0,0)[r]{\strut{}$-5$}}%
\put(1694,7171){\makebox(0,0)[r]{\strut{}$0$}}%
\put(1694,8069){\makebox(0,0)[r]{\strut{}$5$}}%
\end{picture}%
\endgroup
  \end{center}
  \caption{Illustration of a time-frequency selective wireless channel.}
  \label{figure-fft-multipath-rayleigh-channel-envelope}
\end{figure}

In a sparsely scattered channel operated in high-frequency, such as
mmWave, bands, the number of physical paths from transmitter to
receiver may be very small. For this kind of channels, the CIR can be
expressed by \eqref{eq:OTFS-29}, where $h_l$ is a complex number. The
total power conveyed to receiver is
\begin{align}\label{eq:OTFS-32}
P_r=\sum_{l=0}^{\mathcal{L}}\|h_l\|^2
\end{align}
which can be exploited by receiver when it is capable of
distinguishing the individual paths and coherently combines them in
DD-domain. $P_r$ is a constant and invariant unless transmitter,
receiver or/and environment change.

In the TF-domain, as seen in \eqref{eq:OTFS-15} and
\eqref{eq:OTFS-16}, the channel $H[m,n]$ of each TF-element is a
linear combination of the physical propagation paths with different
delays and Doppler-shifts. Hence, $H[m,n]$ is a random process whose
varying rate is dependent on both delays and
Doppler-shifts. Typically, it has the characteristics as shown in
Fig.~\ref{figure-fft-multipath-rayleigh-channel-envelope}. The
TF-elements close to each other experience correlated fading,
but those separated in TF-domain at least by $1/M\Delta f$ or/and
$1/NT$ experience independent fading.

\begin{figure}[tb]
  \begin{center}
    \includegraphics[width=0.9\linewidth]{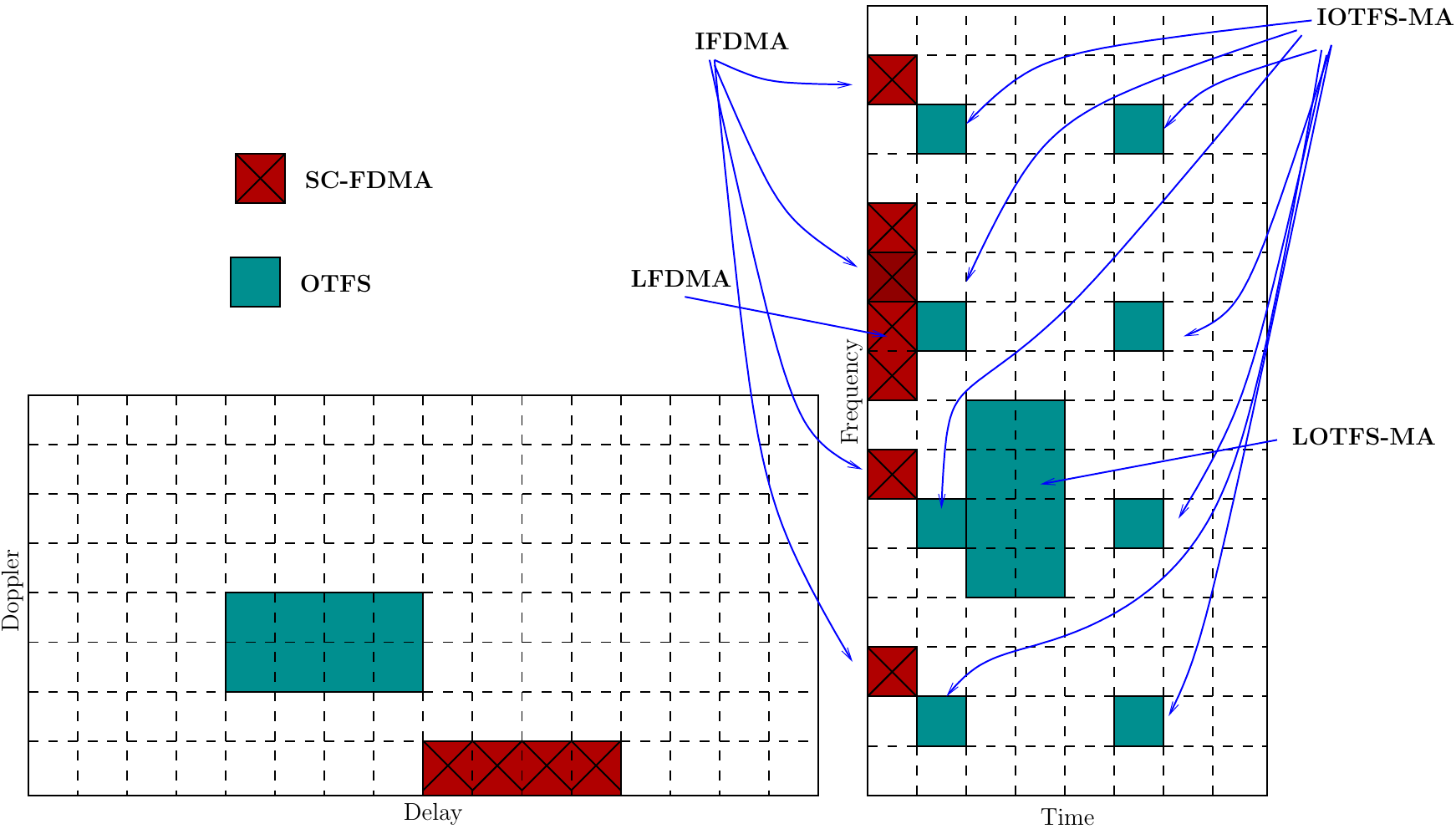}
 
  \end{center}
  \caption{Resource-distribution in OTFS and SC-FDMA multiuser systems.}
\label{figure-OTFS-multiuser}
\end{figure}

Below the multiuser multiplexing is discussed in the preference of
uplink and downlink, respectively, although a scheme may often be
similarly implemented for both uplink and downlink. It is generally
assumed that the transmissions to/from different users are
synchronous. To support the analysis, the DD-domain grid and TF-domain
grid with $16\times 8$ resource units as shown in
Fig.~\ref{figure-OTFS-multiuser} will be referred to. In this example,
we set $M_d=4$ and $N_D=2$, meaning that each user is assigned
$4\times 2$ units of resources in either DD-domain or TF-domain. When
the corresponding SC-FDMA and OFDM are considered, the first row in
DD-domain grid and the first column in TF-domain grid are referred to,
and each user is assigned $M_d=4$ resource units.

\subsection{Uplink}\label{subsection-6G-3.3.1}\index{Uplink|(}

SC-FDMA was proposed mainly for supporting the uplink transmission in
4G LTE/LTE-A systems\index{LTE/LTE-A}. Hence, let us start with the
existing multiplexing (multiple-access) schemes used in SC-FDMA
systems~\cite{Lie-Liang-MC-CDMA-book}. As previously mentioned,
SC-FDMA has two types of mapping protocols from D-domain to F-domain,
yileding IFDMA and LFDMA. Assume that $\tilde{\pmb{x}}_k$ is a
$M_d$-length data symbol vector. For both IFDMA and LFDMA, the signals
mapped to the $M$ subbands (subcarriers) can be written as
\begin{align}\label{eq:OTFS-33}
\pmb{x}_k=\pmb{P}^{(k)}_{M\times M_d}\pmb{\mathcal{F}}_{M_d}\tilde{\pmb{x}}_k
\end{align}
where $\pmb{\mathcal{F}}_{M_d}$ is the $(M_d\times M_d)$ FFT matrix,
$\pmb{P}^{(k)}_{M\times M_d}$ is a $(M\times M_d)$ mapping matrix,
which constitutes $M_d$ columns chosen from $\pmb{I}_M$ according to
the subbands assigned to a user. Specifically for user $k=2$ shown in
the figure, the mapping matrix\index{Mapping!LFDMA} for LFDMA is
\begin{align}\label{eq:OTFS-34}
\setcounter{MaxMatrixCols}{16}
\pmb{P}^T_{16\times 4}=
\begin{bmatrix}
0 & 0 & 0 & 0 & 0 & 0 & 0 & 0 & 1 & 0 & 0 & 0 & 0 & 0 & 0 & 0 \\
0 & 0 & 0 & 0 & 0 & 0 & 0 & 0 & 0 & 1 & 0 & 0 & 0 & 0 & 0 & 0 \\
0 & 0 & 0 & 0 & 0 & 0 & 0 & 0 & 0 & 0 & 1 & 0 & 0 & 0 & 0 & 0 \\
0 & 0 & 0 & 0 & 0 & 0 & 0 & 0 & 0 & 0 & 0 & 1 & 0 & 0 & 0 & 0 
\end{bmatrix}
\end{align}
and for IFDMA\index{Mapping!IFDMA} is 
\begin{align}\label{eq:OTFS-35}
\setcounter{MaxMatrixCols}{16}
\pmb{P}^T_{16\times 4}=
\begin{bmatrix}
0 & 0 & 1 & 0 & 0 & 0 & 0 & 0 & 0 & 0 & 0 & 0 & 0 & 0 & 0 & 0 \\
0 & 0 & 0 & 0 & 0 & 0 & 1 & 0 & 0 & 0 & 0 & 0 & 0 & 0 & 0 & 0 \\
0 & 0 & 0 & 0 & 0 & 0 & 0 & 0 & 0 & 0 & 1 & 0 & 0 & 0 & 0 & 0 \\
0 & 0 & 0 & 0 & 0 & 0 & 0 & 0 & 0 & 0 & 0 & 0 & 0 & 0 & 1 & 0 
\end{bmatrix}
\end{align}
The mapping matrix has the property $\left(\pmb{P}^{(k)}_{M\times
  M_d}\right)^T\pmb{P}^{(k)}_{M\times M_d}=\pmb{I}_{M_d}$.

Generalizing to OTFS, let $\tilde{\pmb{X}}_k$ be the $(N_D\times M_d)$
data symbol block in DD-domain to be transmitted by user $k$. Then,
the signals mapped to TF-domain grid can be expressed as
\begin{align}\label{eq:OTFS-36}
\pmb{X}_k=\pmb{P}^{(k)}_{M\times M_d}\pmb{\mathcal{F}}_{M_d}\tilde{\pmb{X}}_k^T\pmb{\mathcal{F}}_{N_D}^H\pmb{P}^{(k)}_{N_D\times N}
\end{align}
where $\pmb{P}^{(k)}_{N_D\times N}$ is a mapping matrix with its $N_D$
rows chosen from $\pmb{I}_N$ according to the time-slots assigned to
user $k$, which is similar as \eqref{eq:OTFS-34} or
\eqref{eq:OTFS-35}, and has the property of $\pmb{P}^{(k)}_{N_D\times
  M}\left(\pmb{P}^{(k)}_{N_D\times M}\right)^T=\pmb{I}_{N_D}$. Similar
to LFDMA, when the mapping matrices $\pmb{P}^{(k)}_{M\times M_d}$ and
$\pmb{P}^{(k)}_{N_D\times M}$ have the structure of
\eqref{eq:OTFS-34}, the OTFS can be termed as
LOTFS-MA\index{Mapping!LOTFS-MA}.  Similar to IFDMA, when the mapping
matrices $\pmb{P}^{(k)}_{M\times M_d}$ and $\pmb{P}^{(k)}_{N_D\times
  M}$ have the structure of \eqref{eq:OTFS-35}, the OTFS can be termed
as IOTFS-MA\index{Mapping!IOTFS-MA}. In
Fig.~\ref{figure-OTFS-multiuser}, the examples of mappings in LOTFS-MA
and IOTFS-MA are illustrated.

Following IFDMA and LFDMA, it is expected that LOTFS- and IOTFS-MA
have the following properties:
\begin{itemize}

\item As there is a FFT from D-domain to F-domain and an IFFT from
  F-domain to T-domain, as IFDMA/LFDMA, each user's transmit power is
  focused on one frequency, resulting in single-carrier
  transmission. Hence, it contributes to PAPR\index{PAPR}
  reduction. By contrast, there is only an IFFT of
  $\pmb{\mathcal{F}}_{N_D}^H$ from Doppler-domain to T-domain, which
  generates PAPR dynamics. Hence, it can be expected that the transmit
  signals of a user fluctuates in T-domain with respect to
  time-slots. Due to this $N_D$-point IFFT, the PAPR of user's
  transmit signals is $N_D$.

\item IOTFS-MA implements fixed TF-domain resource-allocation, while in
  LOTFS-MA systems, blocks of TF-domain resources may be assigned to
  different users dynamically.

\item Hence, from
  Fig.~\ref{figure-fft-multipath-rayleigh-channel-envelope} we can be
  inferred that IOTFS-MA systems may attain TF-diversity
  gain\index{Diversity ! time-frequency} but no multiuser
  diversity\index{Diversity!multiuser} gain. By contrast, LOTFS-MA
  systems may not gain benefit from channel's TF-selectivity but can
  enjoy multiuser diversity via TF-domain resource-allocation.

\item The maximum diversity order achievable by IOTFS-MA systems is
  limited by $M_dN_D$, and $L_{\max}V_{\max}$ in richly scattered
  channels or $\mathcal{L}$ in sparsely scattered channels. In
  LOTFS-MA systems, the diversity gain is mainly relied on the number
  of users.

\item In IOTFS-MA, when transmitter has channel knowledge, the mapping
  matrices can be replaced by the one obtained from the dynamic
  TF-domain resource-allocation among $K_dK_D$ uplink users. In this
  case, IOTFS-MA is enabled to achieve both TF-diversity and multiuser
  diversity. The cost for this is the possibly increased PAPR.

\end{itemize}

In terms of performance, we should expect the following observations,
when assuming that the ambiguity function satisfies the condition of
\eqref{eq:OTFS-14}.
\begin{itemize}

\item Since there is no interference in TF-domain, as seen in
  \eqref{eq:OTFS-20}, simple one-tap equalization can be implemented
  in TF-domain, and yield no multiuser interference
  (MUI)\index{Multiuser interference}. However, the symbols of one
  user interfere with each other in DD-domain, generating the
  intra-user interference (IUI)\index{Intra-user interference}.

\item Alternatively, equalization can also be implemented in DD-domain
  by first mapping $Y[m,n]$ in \eqref{eq:OTFS-20} without TF-domain
  processing to DD-domain. However, the complexity is higher than that
  operated in TF-domain, as the equalizer is now $M\times N$
  dimensions.

\item While IOTFS-MA enables TF-diversity gain\index{Time-frequency
  diversity} and LOTFS-MA also has the potential to provide certain
  TF-diversity gain, the TF-diversity gain may not be obtained, unless
  IUI is sufficiently suppressed with the aid of an advanced
  equalizer/detector, such as, in the principles of MMSE or maximum
  likelihood.

\end{itemize}

Instead of the mapping implemented in TF-domain, as shown in
\eqref{eq:OTFS-36}, the mapping can also be carried out in
DD-domain. Correspondingly, we have
\begin{align}\label{eq:OTFS-37}
\pmb{X}_k=\pmb{\mathcal{F}}_{M}\pmb{P}^{(k)}_{M\times M_d}\tilde{\pmb{X}}_k^T\pmb{P}^{(k)}_{N_D\times N}\pmb{\mathcal{F}}_{N}^H
\end{align}
In this way, each user's data symbols are firstly mapped to $M_dN_D$
elements of the $(N\times M)$ DD-grid and then a $(M\times N)$-dimensional ISFFT is
executed to transform signals from DD-domain to TF-domain. Comparing
\eqref{eq:OTFS-37} with \eqref{eq:OTFS-36}, which carries out
$(M_d\times N_D)$-dimensional ISFFT, we can be implied that the
transmitter implementation with the scheme of \eqref{eq:OTFS-36} is
less demanding.

Define $\pmb{\mathcal{F}}^{(k)}_{M\times
  M_d}=\pmb{\mathcal{F}}_{M}\pmb{P}^{(k)}_{M\times M_d}$, which in fact is a
$(M\times M_d)$ matrix containing the columns of
$\pmb{\mathcal{F}}_{M}$ selected by $\pmb{P}^{(k)}_{M\times M_d}$. Similarly,
defined $\left(\pmb{\mathcal{F}}^{(k)}_{N\times N_D}\right)^H=\pmb{P}^{(k)}_{N_D\times
  N}\pmb{\mathcal{F}}_{N}^H$, which is a $(N_D\times N)$ matrix
containing the rows of $\pmb{\mathcal{F}}_{N}^H$ chosen by
$\pmb{P}^{(k)}_{N_D\times N}$. Then, \eqref{eq:OTFS-37} is reduced to 
\begin{align}\label{eq:OTFS-38}
\pmb{X}_k=\pmb{\mathcal{F}}^{(k)}_{M\times
  M_d}\tilde{\pmb{X}}_k^T\left(\pmb{\mathcal{F}}^{(k)}_{N\times N_D}\right)^H
\end{align}
Explicitly, the operations can be explained as the 2-D spreading of
$\tilde{\pmb{X}}_k^T$ using the DFT sequences from
$\pmb{\mathcal{F}}_{M}$ and $\pmb{\mathcal{F}}_{N}$. Hence, the scheme can be referred to as the
ISFFT-spread OSTF, corresponding to the DFT-spread OFDM. To generalize
the ISFFT-spread OSTF, \eqref{eq:OTFS-38} can be written as
\begin{align}\label{eq:OTFS-39}
\pmb{X}_k=\pmb{\mathcal{S}}^{(k)}_A\tilde{\pmb{X}}_k^T\left(\pmb{\mathcal{S}}^{(k)}_B\right)^T
\end{align}
where $\pmb{\mathcal{S}}^{(k)}_A$ and $\pmb{\mathcal{S}}^{(k)}_B$ are
general spreading matrices satisfying the requirements for
dimensions. This scheme is referred to as TF-spread OSTF, just like
F-spread CDMA~\cite{Lie-Liang-MC-CDMA-book}.

The characteristics and performance of ISFFT-spread OSTF and TF-spread
OSTF can be summarized as follows.
\begin{itemize}

\item While in ISFFT-spread OSTF, data symbols can be viewed to be
  modulated in DD-domain, in TF-spread OSTF, data symbols are
  modulated and spread in TF-domain. Hence, TF-spread OSTF belongs to
  a TF-domain signaling scheme, with ISFFT-spread OSTF being one of
  its special cases.

\item As each symbol is spread to all the elements of TF-grid, full
  TF-diversity is achievable, when advanced multiuser detectors are
  employed.

\item Fixed resource-allocation (or so-said no resource-allocation) in
  TF-domain and also in power (P)-domain\index{Domain!power (P)},
  except the possible power-control\index{Power-control} for dealing
  with the near-far problem\index{Near-far problem}.

\item Detectors/equalizers need to consider $(M\times N)$ dimensions
  and hence, it may be high-complexity, especially, when advanced
  multiuser detection is implemented.

\item While ISFFT-spread OSTF signals can be expected to have low
  PAPR, TF-spread OSTF signals may experience the PAPR
  problem\index{PAPR problem} depending on the spreading sequences employed.

\end{itemize}

Finally, with the aid of the formula of $\textrm{vec}(\pmb{ABC})=(\pmb{C}^T\otimes\pmb{A})\textrm{vec}(\pmb{B})$, where $\otimes$ denotes the Kronecker product, $\textrm{vec}(\cdot)$ represents the vectoring operation, \eqref{eq:OTFS-39} can be written as
\begin{align}\label{eq:OTFS-40}
\textrm{vec}(\pmb{X}_k)=&(\pmb{\mathcal{S}}^{(k)}_B\otimes \pmb{\mathcal{S}}^{(k)}_A)\textrm{vec}(\tilde{\pmb{X}}_k^T)\\
\label{eq:OTFS-41}
=&\pmb{\mathcal{S}}^{(k)}\times\textrm{vec}(\tilde{\pmb{X}}_k^T)
\end{align}
where $\pmb{\mathcal{S}}^{(k)}=(\pmb{\mathcal{S}}^{(k)}_B\otimes \pmb{\mathcal{S}}^{(k)}_A)$. \eqref{eq:OTFS-40} and \eqref{eq:OTFS-41} explain that the transmitter first spreads a $M_dN_D$-length data vector $\textrm{vec}(\tilde{\pmb{X}}_k^T)$ to obtained a $MN$-length one $\textrm{vec}(\pmb{X}_k)$, the elements of which are then mapped to the $(M\times N)$-grid in TF-domain. 

The difference between \eqref{eq:OTFS-41} and
\eqref{eq:OTFS-40}/\eqref{eq:OTFS-39} is that in
\eqref{eq:OTFS-40}/\eqref{eq:OTFS-39}, the spreading sequences are
obtained by concatenating two sets of sequences of length-$M$ and
length-$N$, respectively, while in \eqref{eq:OTFS-41}, the spreading
sequences are from one generalized set of length-$MN$, which includes
that in \eqref{eq:OTFS-40}/\eqref{eq:OTFS-39} as a subset. Hence,
based on \eqref{eq:OTFS-41}, there are more sequences to choose for
$\pmb{\mathcal{S}}^{(k)}$ and, consequently, yielding better
performance. For example, when binary spreading sequences are
considered, the total number of sequences following
\eqref{eq:OTFS-40}/\eqref{eq:OTFS-39} is $2^{M+N}$, while that
following \eqref{eq:OTFS-41} is $2^{MN}$. Again, we should realized
that in the general cases, the detection complexity may be extreme,
when $MN$ is a big value.

\index{Uplink|)}

\subsection{Downlink}\label{subsection-6G-3.3.2}\index{Downlink|(}

Downlink transmission is in the fashion of broadcast, i.e., BS sends information simultaneously to all downlink users. Hence, BS carries out all transmit processing and the transmissions to different users are synchronous. Below are some possible downlink signaling and transmission schemes. 

Let $\tilde{\pmb{X}}_k\in\mathbb{C}^{N_D\times M_d}$, $k=1,\ldots,K_dK_D$, be the data symbols to be sent to the downlink users. The first downlink signaling scheme maps these data symbols to the $(N\times M)$ grid in DD-domain. Then, they are ISFFT processed to TF-domain. This can be expressed as 
\begin{align}\label{eq:OTFS-42}
\pmb{X}=\pmb{\mathcal{F}}_{M}\left(\sum_{k=1}^{K_dK_D}\pmb{P}^{(k)}_{M\times M_d}\tilde{\pmb{X}}_k^T\pmb{P}^{(k)}_{N_D\times N}\right)\pmb{\mathcal{F}}_{N}^H
\end{align}
where $\pmb{P}^{(k)}_{M\times M_d}$ and $\pmb{P}^{(k)}_{N_D\times N}$
are the mapping matrices assigned to user $k$, which are also used by
user $k$ to pick up its detected information from the
DD-grid. Equation \eqref{eq:OTFS-42} can be written as
\begin{align}\label{eq:OTFS-43}
\pmb{X}=&\sum_{k=1}^{K_dK_D}\pmb{\mathcal{F}}_{M}\pmb{P}^{(k)}_{M\times M_d}\tilde{\pmb{X}}_k^T\pmb{P}^{(k)}_{N_D\times N}\pmb{\mathcal{F}}_{N}^H\nonumber\\
=& \sum_{k=1}^{K_dK_D}\pmb{\mathcal{F}}^{(k)}_{M\times M_d}\tilde{\pmb{X}}_k^T\left(\pmb{\mathcal{F}}^{(k)}_{N\times N_D}\right)^H
\end{align}
where $\pmb{\mathcal{F}}^{(k)}_{M\times M_d}$ consists of the $M_d$
columns/sequences of $\pmb{\mathcal{F}}_{M}$ while
$\pmb{\mathcal{F}}^{(k)}_{N\times N_D}$ consists of the $N_D$
columns/sequences of $\pmb{\mathcal{F}}_{N}$ assigned to user
$k$. Hence, with this downlink signaling scheme, each symbol is spread
to all the $MN$ resource units in TF-domain. Therefore, while each
user can enjoy the TF-diversity, if it can afford an equalizer of
relatively high-complexity, there is no multiuser diversity available
from the resource-allocation in TF-domain. Only power-control is
needed in response to the transmission distances between users and
BS.

Secondly, as \eqref{eq:OTFS-39}, the above signaling scheme can be extended to a downlink TF-spread OSTF scheme, which accordingly, has the formula
\begin{align}\label{eq:OTFS-44}
\pmb{X}=& \sum_{k=1}^{K_dK_D}\pmb{\mathcal{S}}^{(k)}_A\tilde{\pmb{X}}_k^T\left(\pmb{\mathcal{S}}^{(k)}_B\right)^T
\end{align}
Furthermore, following \eqref{eq:OTFS-41}, the most general downlink TF-spread scheme generates the TF-domain signals as
\begin{align}\label{eq:OTFS-45}
\textrm{vec}(\pmb{X})=& \sum_{k=1}^{K_dK_D}\pmb{\mathcal{S}}^{(k)}\times\textrm{vec}(\tilde{\pmb{X}}_k^T)
\end{align}
Again, it is worth noting that these spreading schemes cannot gain
benefit from the TF-domain resource-allocation to achieve multiuser
diversity.

However, it is noteworthy that $\pmb{\mathcal{S}}^{(k)}$ in
\eqref{eq:OTFS-45} is very general and can be configured for designing
different signaling schemes. For example, $\pmb{\mathcal{S}}^{(k)}$
can be designed to interleave the elements in $\tilde{\pmb{X}}_k$ and
different users are distinguished by their interleaving matrices. It
can be a sparse matrix to implement sparse code-division
multiple-access (SCDMA)~\cite{8166802,Sparse-4471881} or sparse-code
multiple-access (SCMA)~\cite{7263349,9411900}. Or, it is a matrix
chosen from a set, which is joined with $\tilde{\pmb{X}}_k$ to
implement TF-domain index modulation (IM)\index{Index modulation
  (IM)}~\cite{10172295,10129061}. It is easy to understand that the
indices can be defined on the activities of the elements in
$\tilde{\pmb{X}}_k$, or the codes in
$\pmb{\mathcal{S}}^{(k)}$. Furthermore, a set of matrices of
$\{\pmb{\mathcal{S}}^{(k)}\}$ can be designed as modulation indices
for each user, and IM is implemented at matrix level. Additionally,
when multiple transmit and receive antennas are employed, spatial
modulation (SM)\index{Spatial modulation (SM)}~\cite{10079919} based
OSTF or preprocessing-SM (PSM)\index{Preprocessing-assisted
  spatial modulation (PSM)} based OSTF~\cite{5956573} can be
designed.

The above downlink signaling schemes are designed by assuming that CSI
is employed at user receivers, BS only knows the statistical CSI to
each of users. When BS also employs the CSI to all users, the
spreading in \eqref{eq:OTFS-45} can be replaced by
preprocessing\index{Preprocessing} as
\begin{align}\label{eq:OTFS-46}
\textrm{vec}(\pmb{X})=& \sum_{k=1}^{K_dK_D}\pmb{\mathcal{P}}^{(k)}\times\textrm{vec}(\tilde{\pmb{X}}_k^T)\nonumber\\
=&\pmb{\mathcal{P}}\tilde{\pmb{x}}
\end{align}
where 
\begin{align}\label{eq:OTFS-47}
\pmb{\mathcal{P}}=&\left[\pmb{\mathcal{P}}^{(1)},\pmb{\mathcal{P}}^{(2)},\ldots,\pmb{\mathcal{P}}^{(K_dK_D)}\right]\nonumber\\
 \tilde{\pmb{x}}=&\left[\textrm{vec}^T(\tilde{\pmb{X}}_1^T),\textrm{vec}^T(\tilde{\pmb{X}}_2^T),\ldots,\textrm{vec}^T(\tilde{\pmb{X}}_{K_dK_D}^T)\right]^T
\end{align}
with $\textrm{vec}^T(\cdot)$ denoting the transpose operation of a vector obtained from vectorization. This gives the third downlink transmission scheme, which carries out transmit preprocessing in TF-domain.

Since in \eqref{eq:OTFS-46}, $\pmb{\mathcal{P}}$ is a transmit
preprocessing matrix need to be optimized with certain objective under
required constraints by exploiting the CSI to different users, the
resource-allocation in both TF- and P-domains have to be
addressed. For example, BS may first make use the TF-domain DoFs to
design a preprocessing matrix $\pmb{\mathcal{P}}'$, so that the
signals arriving at different users are free of MUI, enabling mobile
users to implement low-complexity detection. Then, power is allocated
to obtain $\pmb{\mathcal{P}}=\pmb{\mathcal{P}}'\pmb{\beta}$, where
$\pmb{\beta}$ is a real diagonal matrix for power-allocation to
maximize SNR, spectral-efficiency, etc., of the
system~\cite{Lie-Liang-MC-CDMA-book}.

Following OFDM, the final downlink transmission scheme considered can be described as
\begin{align}\label{eq:OTFS-48}
\pmb{X}=&\sum_{k=1}^{K_dK_D}\pmb{P}^{(k)}_{M\times M_d}\tilde{\pmb{X}}_k^T\pmb{\beta}_k\pmb{P}^{(k)}_{N_D\times N}
\end{align}
This is a pure resource-allocation scheme operated in TF-domain, i.e.,
it is an OSTF scheme. Depended on the expected trade-off between
efficiency and complexity, $\pmb{P}^{(k)}_{M\times M_d}$,
$\pmb{P}^{(k)}_{N_D\times N}$ and $\pmb{\beta}_k$ may be designed as
follows.  Firstly, it may do fixed allocation with
$\pmb{P}^{(k)}_{M\times M_d}$, $\pmb{P}^{(k)}_{N_D\times N}$ and
$\pmb{\beta}_k$ (power-allocation matrix) all fixed for the individual
users. Secondly, it may fix the TF-domain resources using fixed
$\pmb{P}^{(k)}_{M\times M_d}$ and $\pmb{P}^{(k)}_{N_D\times N}$ for
individual users, but allocate power, i.e., design $\pmb{\beta}_k$, to
users dynamically. Lastly, it can dynamically allocate both TF-domain
resources and power via designing $\pmb{P}^{(k)}_{M\times M_d}$,
$\pmb{P}^{(k)}_{N_D\times N}$ and $\pmb{\beta}_k$ according to the CSI
to different users, service requirements, and so on. It may implement
joint optimization, or carry out TF- and P-domains\index{Domain!power
  (P)} resource-allocation separately to reduce complexity. For
example, BS may first assign each downlink user the TF-domain
resources (subbands and time-slots), followed by the power-allocation
in the principle of, such as, water-filling.  With this regard, the
optimization algorithms can be straightforwardly extended from that
provided in Chapters~3 and 4.

\index{Downlink|)}
\section{Concluding Remarks}\label{subsection-6G-3.4}

The principles of OTFS and its companion OSTF have been analyzed,
showing that they represent the 1D-to-2D\index{1D: One
  dimension}\index{2D: two dimensions} extensions of SC-FDMA and OFDM,
respectively. As SC-FDMA and OFDM, we can be inferred that OTFS and
OSTF have their individual advantages and disadvantages that are
dependent on application scenarios. Typically, as SC-FDMA able to
achieve F-diversity, OTFS is beneficial to obtaining TF-diversity
resulted from the delay- and Doppler-spread, leading any DD-domain
symbols to attain similar performance. Due to this, OTFS signaling is
not feasible for exploiting the DoFs in space
(S)-domain\index{Domain!space (S)}, provided by distributed users, to
obtain multiuser diversity. By contrast, as OFDM, single TF-domain
symbol in OSTF cannot benefit from channel's frequency- and
time-selectivity to get diversity gain. However, the qualities of
different TF-domain elements in OSTF can be highly dynamic, which
renders resource-allocation to be high-efficiency. OSTF signals in
TF-domain are highly time-variant, but have little interference with
each other. In contrast, OTFS signals in DD-domain are slowly
time-varying signals, but they interfere with each other.

The practical implementations of both OTFS and OSTF face many
challenges, but most of them are rooted from channel estimation, i.e.,
estimating $h(\tau,\nu)$ as accurate as possible and with an overhead as
small as possible. Provided that accurate $h(\tau,\nu)$ is available,
mapping data in DD-domain or alternatively, in TF-domain is not a
critical issue, but mainly dependent on the feasibility of
implementation, which is limited by practical communications
environments. From the principles of OTFS and OSTF, we can be inferred
that for channel estimation, pilot symbols may be arranged in
DD-domain~\cite{hadani2018orthogonal,8727425,8671740,10151793} or in
TF-domain~\cite{10151793}, and correspondingly, channels can be
estimated in DD-domain or TF-domain. However, it is noteworthy that,
regardless of OTFS or OSTF, channels can be estimated either in
DD-domain or in TF-domain. This is analogous to SC-FDMA and OFDM, no
matter which one it is, the objective of channel estimation is to
estimate the CIR in D-domain, which can be implemented either in
F-domain or in D-domain. The difference between them is that in
D-domain, the estimator aims to recover the accurate $h(\tau)$ by
estimating all individual rays. Hence, it is more suitable for
estimating the sparse channels having a low number of rays from
transmitter to receiver. By contrast, in F-domain, the estimator
motivates to find the $h(\tau)$ that can reveal the channel behaviors
on different subcarriers, but does not put emphasis on whether the
estimated $h(\tau)$ is the true CIR. This is especially suitable for
estimating the channels having many multipath components generated by
random scatters.  Again, which one is more desirable is dependent on
the communication environments in practice and other related factors,
such as, complexity-efficiency trade-off.


\begin{thebibliography}{10}
\providecommand{\url}[1]{#1}
\csname url@samestyle\endcsname
\providecommand{\newblock}{\relax}
\providecommand{\bibinfo}[2]{#2}
\providecommand{\BIBentrySTDinterwordspacing}{\spaceskip=0pt\relax}
\providecommand{\BIBentryALTinterwordstretchfactor}{4}
\providecommand{\BIBentryALTinterwordspacing}{\spaceskip=\fontdimen2\font plus
\BIBentryALTinterwordstretchfactor\fontdimen3\font minus
  \fontdimen4\font\relax}
\providecommand{\BIBforeignlanguage}[2]{{%
\expandafter\ifx\csname l@#1\endcsname\relax
\typeout{** WARNING: IEEEtran.bst: No hyphenation pattern has been}%
\typeout{** loaded for the language `#1'. Using the pattern for}%
\typeout{** the default language instead.}%
\else
\language=\csname l@#1\endcsname
\fi
#2}}
\providecommand{\BIBdecl}{\relax}
\BIBdecl

\bibitem{9685566}
A.~M. Sayeed, ``How is time frequency space modulation related to short time
  fourier signaling?'' in \emph{2021 IEEE Global Communications Conference
  (GLOBECOM)}, 2021, pp. 1--6.

\bibitem{1347350}
K.~Liu, T.~Kadous, and A.~Sayeed, ``Orthogonal time-frequency signaling over
  doubly dispersive channels,'' \emph{IEEE Transactions on Information Theory},
  vol.~50, no.~11, pp. 2583--2603, 2004.

\bibitem{7925924}
R.~Hadani, S.~Rakib, M.~Tsatsanis, A.~Monk, A.~J. Goldsmith, A.~F. Molisch, and
  R.~Calderbank, ``Orthogonal time frequency space modulation,'' in \emph{2017
  IEEE Wireless Communications and Networking Conference (WCNC)}, 2017, pp.
  1--6.

\bibitem{hadani2018orthogonal}
R.~Hadani, S.~Rakib, S.~Kons, M.~Tsatsanis, A.~Monk, C.~Ibars, J.~Delfeld,
  Y.~Hebron, A.~J. Goldsmith, A.~F. Molisch, and R.~Calderbank, ``Orthogonal
  time frequency space modulation,'' 2018.

\bibitem{8424569}
P.~Raviteja, K.~T. Phan, Y.~Hong, and E.~Viterbo, ``Interference cancellation
  and iterative detection for orthogonal time frequency space modulation,''
  \emph{IEEE Transactions on Wireless Communications}, vol.~17, no.~10, pp.
  6501--6515, 2018.

\bibitem{COM-Bello-1963}
P.~Bello, ``Characterization of randomly time-variant linear channels,''
  \emph{IEEE Transactions on Communications}, vol.~11, no.~4, pp. 360 -- 393,
  December 1963.

\bibitem{9321356}
Z.~Wei, W.~Yuan, S.~Li, J.~Yuan, and D.~W.~K. Ng, ``Transmitter and receiver
  window designs for orthogonal time-frequency space modulation,'' \emph{IEEE
  Transactions on Communications}, vol.~69, no.~4, pp. 2207--2223, 2021.

\bibitem{Sayeed-TF-perspective}
A.~M. Sayeed and B.~Aazhang, ``Communication over multipath fading channels:
  {A} time-frequency perspective,'' in \emph{Wireless Communication - {TDMA}
  versus {CDMA}}, S.~G. Glisic and P.~L. Leppanen, Eds.\hskip 1em plus 0.5em
  minus 0.4em\relax Kluwer Academic Publishers, 1997, pp. 73--98.

\bibitem{Cohen-Time-Frequency-book}
L.~Cohen, \emph{Time-Frequency Analysis}.\hskip 1em plus 0.5em minus
  0.4em\relax Englewood Cliffs, New Jersey, USA: Prentice Hall PTR, 1995.

\bibitem{book:Zhang-Xianda}
X.~Zhang, \emph{Matrix Analysis and Applications}.\hskip 1em plus 0.5em minus
  0.4em\relax Beijing, P.R.China: Tsinghua University Press, 2006.

\bibitem{Lie-Liang-MC-CDMA-book}
L.-L. Yang, \emph{Multicarrier Communications}.\hskip 1em plus 0.5em minus
  0.4em\relax Chichester, United Kingdom: John Wiley, 2009.

\bibitem{J-Adachi-IEICE-2009-1441}
F.~Adachi, H.~Tomeba, and K.~Takeda, ``Frequency-domain equalization for
  broadband single-carrier multiple access,'' \emph{IEICE Transactions on
  Communications}, vol. E92-B, no.~5, pp. 1441 -- 1456, May 2009.

\bibitem{5493922}
S.~Okuyama, K.~Takeda, and F.~Adachi, ``{MMSE} frequency-domain equalization
  using spectrum combining for {Nyquist} filtered broadband single-carrier
  transmission,'' in \emph{2010 IEEE 71st Vehicular Technology Conference},
  2010, pp. 1--5.

\bibitem{Proakis-5th}
J.~G. Proakis, \emph{Digital Communications}, 5th~ed.\hskip 1em plus 0.5em
  minus 0.4em\relax McGraw Hill, 2007.

\bibitem{8166802}
Y.~Liu, L.-L. Yang, and L.~Hanzo, ``Spatial modulation aided sparse
  code-division multiple access,'' \emph{IEEE Transactions on Wireless
  Communications}, vol.~17, no.~3, pp. 1474--1487, 2018.

\bibitem{Sparse-4471881}
R.~Hoshyar, F.~Wathan, and R.~Tafazolli, ``Novel low-density signature for
  synchronous {CDMA} systems over {AWGN} channel,'' \emph{IEEE Transactions on
  Signal Processing}, vol.~56, no.~4, pp. 1616--1626, April 2008.

\bibitem{7263349}
L.~Dai, B.~Wang, Y.~Yuan, S.~Han, I.~Chih-lin, and Z.~Wang, ``Non-orthogonal
  multiple access for {5G}: solutions, challenges, opportunities, and future
  research trends,'' \emph{IEEE Communications Magazine}, vol.~53, no.~9, pp.
  74--81, 2015.

\bibitem{9411900}
K.~Deka, A.~Thomas, and S.~Sharma, ``{OTFS-SCMA}: A code-domain {NOMA} approach
  for orthogonal time frequency space modulation,'' \emph{IEEE Transactions on
  Communications}, vol.~69, no.~8, pp. 5043--5058, 2021.

\bibitem{10172295}
S.~Doğan-Tusha, A.~Tusha, S.~Althunibat, and K.~Qaraqe, ``Orthogonal time
  frequency space multiple access using index modulation,'' \emph{IEEE
  Transactions on Vehicular Technology}, vol.~72, no.~12, pp. 15\,858--15\,866,
  2023.

\bibitem{10129061}
Z.~Sui, H.~Zhang, Y.~Xin, T.~Bao, L.-L. Yang, and L.~Hanzo, ``Low complexity
  detection of spatial modulation aided {OTFS} in doubly-selective channels,''
  \emph{IEEE Transactions on Vehicular Technology}, vol.~72, no.~10, pp.
  13\,746--13\,751, 2023.

\bibitem{10079919}
Y.~Yang, Z.~Bai, K.~Pang, S.~Guo, H.~Zhang, and K.~S. Kwak, ``Spatial-index
  modulation based orthogonal time frequency space system in vehicular
  networks,'' \emph{IEEE Transactions on Intelligent Transportation Systems},
  vol.~24, no.~6, pp. 6165--6177, 2023.

\bibitem{5956573}
L.-L. Yang, ``Transmitter preprocessing aided spatial modulation for
  multiple-input multiple-output systems,'' in \emph{2011 IEEE 73rd Vehicular
  Technology Conference (VTC Spring)}, 2011, pp. 1--5.

\bibitem{8727425}
W.~Shen, L.~Dai, J.~An, P.~Fan, and R.~W. Heath, ``Channel estimation for
  orthogonal time frequency space {(OTFS)} massive {MIMO},'' \emph{IEEE
  Transactions on Signal Processing}, vol.~67, no.~16, pp. 4204--4217, 2019.

\bibitem{8671740}
P.~Raviteja, K.~T. Phan, and Y.~Hong, ``Embedded pilot-aided channel estimation
  for {OTFS} in delay–doppler channels,'' \emph{IEEE Transactions on
  Vehicular Technology}, vol.~68, no.~5, pp. 4906--4917, 2019.

\bibitem{10151793}
H.-T. Sheng and W.-R. Wu, ``Time-frequency domain channel estimation for {OTFS}
  systems,'' \emph{IEEE Transactions on Wireless Communications}, pp. 1--1,
  2023.

\end{thebibliography}

\end{document}